\documentclass[aps,prl,twocolumn,superscriptaddress,10pt]{revtex4-2}

\usepackage{graphicx}
\usepackage{dcolumn}
\usepackage{bm}
\usepackage[colorlinks,linkcolor=blue,anchorcolor=blue,citecolor=blue,urlcolor=blue]{hyperref}
\usepackage{epstopdf}
\usepackage{textcomp}
\usepackage{gensymb}
\usepackage{booktabs}
\usepackage{orcidlink}
\usepackage{mwe}
\usepackage{color}
\usepackage{url}
\usepackage{siunitx}
\usepackage{subfigure}
\usepackage{longtable}
\usepackage{array}
\usepackage{verbatim}
\usepackage{multirow}
\usepackage{dcolumn}
\usepackage{graphicx}
\usepackage{hyperref}
\usepackage{color}
\usepackage[english]{babel}
\usepackage{bm}
\usepackage{float}
\usepackage{threeparttable}
\usepackage{ulem}
\emergencystretch=\maxdimen
\usepackage{dcolumn}
\newcolumntype{x}{D{.}{.}{6.6}}
\newcolumntype{y}{D{.}{.}{4.5}}
\newcolumntype{z}{D{.}{.}{5.7}}
\newcolumntype{f}{D{.}{.}{7.9}}
\newcolumntype{e}{D{.}{.}{5.6}}

\begin{document}

\preprint{APS/123-QED}

\title{Shell Evolution and Cross-Shell Correlations near $^{78}$Ni Revealed by the Spin and Electromagnetic Moments of $^{81}$Zn}

\author{Y.~C.~Liu}
\affiliation{School of Physics and State Key Laboratory of Nuclear Physics and Technology, Peking University, Beijing 100871, China}

\author{X.~F.~Yang\orcidlink{0000-0002-1633-4000}}
\email{xiaofei.yang@pku.edu.cn}
\affiliation{School of Physics and State Key Laboratory of Nuclear Physics and Technology, Peking University, Beijing 100871, China}

\author{L.~A.~Lalanne}
\altaffiliation[Present address: ]{Université Paris-Saclay, CNRS/IN2P3, IJCLab, Orsay, 91405, France}
\affiliation{Experimental Physics Department, CERN, CH-1211 Geneva 23, Switzerland}

\author{\mbox{M.~Athanasakis-Kaklamanakis\orcidlink{0000-0003-0336-5980}}}	
\altaffiliation[Present address: ]{JILA and University of Colorado, Boulder, CO 80309, USA}
\affiliation{Experimental Physics Department, CERN, CH-1211 Geneva 23, Switzerland}
\affiliation{KU Leuven, Instituut voor Kern- en Stralingsfysica, B-3001 Leuven, Belgium}

\author{\mbox{M.~Au}}	
\affiliation{Systems Department, CERN, CH-1211 Geneva 23, Switzerland}
\affiliation{Johannes Gutenberg-Universit{\"a}t Mainz, 55099 Mainz, Germany}

\author{S.~W.~Bai}
\affiliation{School of Physics and State Key Laboratory of Nuclear Physics and Technology, Peking University, Beijing 100871, China}
\affiliation{State Key Laboratory of Nuclear Physics and Technology, Institute of Quantum Matter, South China Normal University, Guangzhou 510006, China}

\author{S.~Bara}
\altaffiliation[Present address: ]{Université de Caen Normandie, ENSICAEN, CNRS/IN2P3, LPC Caen, F-14000 Caen, France}
\affiliation{KU Leuven, Instituut voor Kern- en Stralingsfysica, B-3001 Leuven, Belgium}

\author{\mbox{T.~E.~Cocolios}}
\affiliation{KU Leuven, Instituut voor Kern- en Stralingsfysica, B-3001 Leuven, Belgium}

\author{N. ~Chen}
\affiliation{State Key Laboratory of Heavy Ion Science and Technology, Institute of Modern Physics, Chinese Academy of Sciences, Lanzhou 730000, China}

\author{R.~P.~de~Groote}
\affiliation{KU Leuven, Instituut voor Kern- en Stralingsfysica, B-3001 Leuven, Belgium}

\author{K.~T.~Flanagan}
\affiliation{Department of Physics and Astronomy, The University of Manchester, Manchester, M13 9PL, United Kingdom}
\affiliation{Photon Science Institute, Alan Turing Building, The University of Manchester, Manchester M13 9PY, United Kingdom}

\author{R.~F.~Garcia Ruiz\orcidlink{0000-0002-2926-5569}}
\affiliation{Massachusetts Institute of Technology, Cambridge, MA 02139, USA}

\author{Y.~F.~Guo}
\affiliation{School of Physics and State Key Laboratory of Nuclear Physics and Technology, Peking University, Beijing 100871, China}

\author{M.~Heines}
\altaffiliation[Present address: ]{Paul Scherrer Institut, Villigen PSI 5232, Switzerland}
\affiliation{KU Leuven, Instituut voor Kern- en Stralingsfysica, B-3001 Leuven, Belgium}

\author{J.~D.~Holt\orcidlink{0000-0003-4833-7959}}
\affiliation{TRIUMF, Vancouver, British Columbia, BC V6T 2A3, Canada}
\affiliation{Department of Physics, McGill University, Montr\'eal, Quebec, QC H3A 2T8, Canada}

\author{B.~S.~Hu\orcidlink{0000-0001-8071-158X}}
\affiliation{School of Physics and State Key Laboratory of Nuclear Physics and Technology, Peking University, Beijing 100871, China}
\affiliation{Southern Center for Nuclear-Science Theory (SCNT), Institute of Modern Physics, Chinese Academy of Sciences, Huizhou 516000, China}

\author{H.~R.~Hu}
\affiliation{School of Physics and State Key Laboratory of Nuclear Physics and Technology, Peking University, Beijing 100871, China}

\author{\'{A}.~Koszor\'{u}s\orcidlink{0000-0001-7959-8786}}
\affiliation{KU Leuven, Instituut voor Kern- en Stralingsfysica, B-3001 Leuven, Belgium}

\author{P.~Lassegues}
\altaffiliation[Present address: ]{Laboratoire National des Champs Magnétiques Intenses, Toulouse, FRANCE}
\affiliation{KU Leuven, Instituut voor Kern- en Stralingsfysica, B-3001 Leuven, Belgium}

\author{J. G. ~Li}
\affiliation{State Key Laboratory of Heavy Ion Science and Technology, Institute of Modern Physics, Chinese Academy of Sciences, Lanzhou 730000, China}
\affiliation{Southern Center for Nuclear-Science Theory (SCNT), Institute of Modern Physics, Chinese Academy of Sciences, Huizhou 516000, Guangdong Province, China}

\author{R.~Lica\orcidlink{0000-0001-7066-9787}}
\affiliation{``Horia Hulubei'' National Institute for R\& D in Physics and Nuclear Engineering, Bucharest, RO-077125, Romania}

\author{Y.~S.~Liu}
\affiliation{School of Physics and State Key Laboratory of Nuclear Physics and Technology, Peking University, Beijing 100871, China}

\author{K.~M.~Lynch\orcidlink{0000-0001-8591-2700}}
\affiliation{Department of Physics and Astronomy, The University of Manchester, Manchester, M13 9PL, United Kingdom}

\author{T.~Miyagi} 
\affiliation{Center for Computational Sciences, University of Tsukuba, 1-1-1 Tennodai, Tsukuba 305-8577, Japan}
\affiliation{Institut f\"{u}r Kernphysik, Technische Universität Darmstadt, D-64289 Darmstadt, Germany}
\affiliation{ExtreMe Matter Institute EMMI, GSI Helmholtzzentrum für Schwerionenforschung GmbH, 64291 Darmstadt, Germany}
\affiliation{Max-Planck-Institut f\"{u}r Kernphysik, D-69117 Heidelberg, Germany}

\author{A.~McGlone}
\affiliation{Department of Physics and Astronomy, The University of Manchester, Manchester, M13 9PL, United Kingdom}

\author{C.~Neascu}
\affiliation{``Horia Hulubei'' National Institute for R\& D in Physics and Nuclear Engineering, Bucharest, RO-077125, Romania}

\author{G.~Neyens\orcidlink{0000-0001-8613-1455}}
\affiliation{KU Leuven, Instituut voor Kern- en Stralingsfysica, B-3001 Leuven, Belgium}

\author{J.~R.~Reilly}
\affiliation{Department of Physics and Astronomy, The University of Manchester, Manchester, M13 9PL, United Kingdom}

\author{C.~M.~Steenkamp}
\affiliation{Stellenbosch Photonics Institute, Physics Department, Stellenbosch University, 7600, Stellenbosch, South Africa}

\author{\mbox{S.~Stegemann}}	
\affiliation{Systems Department, CERN, CH-1211 Geneva 23, Switzerland}

\author{B.~van den Borne}
\altaffiliation[Present address: ]{CERN, CH-1211 Geneva 23, Switzerland}
\affiliation{KU Leuven, Instituut voor Kern- en Stralingsfysica, B-3001 Leuven, Belgium}

\author{\mbox{J.~Wessolek}}	
\affiliation{Systems Department, CERN, CH-1211 Geneva 23, Switzerland}
\affiliation{Department of Physics and Astronomy, The University of Manchester, Manchester, M13 9PL, United Kingdom}

\author{Z.~Yan}
\affiliation{School of Physics and State Key Laboratory of Nuclear Physics and Technology, Peking University, Beijing 100871, China}

\date{\today}

\begin{abstract}

Collinear resonance ionization spectroscopy has been performed on neutron-rich $^{81}$Zn in the vicinity of doubly-magic $^{78}$Ni. The ground state spin and parity are unambiguously assigned as $I^\pi=5/2^+$, excluding the previously suggested $1/2^+$--$5/2^+$ inversion in $^{81}$Zn. The magnetic dipole moment, $\mu=-0.9486(6)$$\mu_{\rm N}$, and electric quadrupole moment, $Q_s=-0.28(3)$ b, are also determined for the first time. Together with the systematics of the $N=51$ isotones, these moments show increasing deviations from a simple single-particle picture toward the $Z=28$ shell closure. Comparisons with large-scale shell-model and \textit{ab initio} calculations indicate that excitations across the $Z=28$ and $N=50$ shell closures are essential for describing the $^{81}$Zn moments, revealing substantial many-body correlations in the immediate vicinity of doubly-magic $^{78}$Ni.

\end{abstract}

\maketitle


Atomic nuclei are quantum many-body systems in which protons and neutrons arrange themselves under the strong nucleon-nucleon interaction. Despite the complexity of the underlying nuclear force, a remarkably simple pattern emerges: nuclei with certain numbers of protons or neutrons are especially stable. These so-called magic numbers, $Z,N=2,8,20,28,50,82$ and $N=126$, are cornerstones of the nuclear shell model~(SM) and have guided our understanding of nuclear structure for decades. However, this pattern is not universal. In nuclei with extreme proton-to-neutron ratios, far from the valley of $\beta$-stability, nuclear shell gaps may weaken, disappear, or be replaced by new ones~\cite{PhysofExoticNuclei,ShellEvolution-Ostuka,NewMagicNumber-Sorlin}. The shell evolution toward the limits of nuclear existence is therefore one of the central questions in modern nuclear physics~\cite{PhysofExoticNuclei}.

The neutron-rich nucleus $^{78}$Ni lies at the center of this question. With $Z=28$ and $N=50$, it combines two traditional magic numbers under extreme neutron excess. Located six $\beta$ decays away from stable $^{78}$Se, $^{78}$Ni has long been predicted to be the most neutron-rich, doubly-magic nucleus potentially accessible experimentally, and is also an important waiting point in the astrophysical rapid-neutron-capture process~\cite{WaitingPoint-2}. The structure of $^{78}$Ni and its neighboring nuclei therefore provides a stringent benchmark for testing the persistence of magicity, the onset of collectivity, and the microscopic mechanisms driving shell evolution under extreme $N/Z$ ratios.

Direct experimental information on $^{78}$Ni itself, however, remains limited by its low production yield, with only one direct experimental observation using in-beam $\gamma$-ray spectroscopy at RIKEN-RIBF~\cite{Taniuchi2019-Ni}. Neighboring nuclei have therefore played a crucial role as probes of exotic structure around $Z=28$ and $N=50$~\cite{Flanagan2009-Cu,Cheal2010-Ga81,deGroote2017-PRC-Cu,Yang2016-Zn79,Zn-moment,Zn-radii,Cu-radii2020, Taniuchi2019-Ni,SM78Ni,Nies2023-Zn79m,Welker2017-Cu,Olivier2017-Cu79}. In $^{79}$Zn, laser spectroscopy revealed evidence for shape coexistence through the large isomer shift and magnetic moment of a long-lived $1/2^+$ isomer~\cite{Yang2016-Zn79}, an interpretation further supported by SM predictions of a low-lying deformed $0_2^+$ state in $^{78}$Ni~\cite{SM78Ni} and by a high-precision mass measurement of $^{79m}$Zn~\cite{Nies2023-Zn79m}. These intruder configurations so close to $^{78}$Ni appear, at first sight, to challenge a simple picture of $^{78}$Ni as a robust, doubly-magic core. In contrast, binding-energy and $\gamma$-ray spectroscopy measurements of $^{79}$Cu strongly support the persistence of the $Z=28$ and $N=50$ shell closures~\cite{Welker2017-Cu,Olivier2017-Cu79}. This tension was clarified by direct in-beam $\gamma$-ray spectroscopy measurement of $^{78}$Ni, which established a high-lying first $2^+$ state, confirming the doubly magic character of its ground state, while also identifying a low-lying deformed $0_2^+$ state~\cite{Taniuchi2019-Ni}. The same work further suggested a weakening of the $Z=28$ shell closure above $N=50$, favoring deformed ground states in more neutron-rich Ni isotopes. Thus, $^{78}$Ni emerges not as a simple inert core, but as a system where magicity and low-lying deformation coexist.

A key question is then how rapidly the structure changes in the $Z=28$ region as neutrons are added beyond the $N=50$ shell closure. High-resolution laser spectroscopy is well suited to address this question, as it provides direct access to ground state spins and electromagnetic properties, which are sensitive to the single-particle configurations, configuration mixing, and deformation~\cite{PPNP2023}. Since the neutron-rich nickel isotopes are not yet accessible for such measurements~\cite{Kaufmann2020-Ni,Ni-radii}, the neighboring $N=51$ isotones with one neutron outside the $N=50$ shell closure offer a feasible alternative.

Experimentally, a $5/2^+$ ground state is established for $N=51$ isotones from $Z=40$ down to $Z=32$, while the excitation energy of the low-lying $1/2^+$ state decreases rapidly in energy toward lower $Z$~\cite{expN51,81Zngs}. Earlier SM calculations predicted an inversion of the $\nu 2d_{5/2}$ and $\nu 3s_{1/2}$ states driven by tensor-induced monopole shifts in the proton-neutron interaction, yielding a $1/2^+$ ground state in both $^{81}$Zn and $^{79}$Ni~\cite{2009N51,SM-51N}. \textit{Ab initio} coupled-cluster~(CC) calculations predict a similar inversion in $^{79}$Ni attributed instead to continuum effects~\cite{Gaute-78Ni}. The ground state spin of $^{81}$Zn, only three valence nucleons away from $^{78}$Ni and the closest accessible even-$Z$, $N=51$ nucleus to $^{79}$Ni, therefore would provide a decisive test of the predicted shell inversion near $Z=30$. Despite several $\beta$-decay spectroscopy studies over almost two decades~\cite{beta-decay-1,beta-decay-2,beta-decay-3}, the ground state spin of $^{81}$Zn remains undetermined. Moreover, the magnetic dipole and electric quadrupole moments of $^{81}$Zn would provide complementary constraints on the underlying wave function, probing the potential single-particle structure, configuration mixing and deformation. Deviations from the moments expected for a single neutron coupled to an inert doubly-magic $^{78}$Ni core would signal correlations across the $Z=28$ and $N=50$ shell closures.

In this letter, we report the first direct measurement of the ground state properties of neutron-rich $^{81}$Zn using collinear resonance ionization spectroscopy~(CRIS). Its spin and parity are unambiguously determined to be $I^\pi=5/2^+$, excluding a $1/2^+$ ground state and thus disfavoring the predicted inversion of the $5/2^+$ and $1/2^+$ states at $Z=30$. The measured magnetic dipole and electric quadrupole moments, confronted with large-scale shell model~(LSSM) and \textit{ab initio} valence-space in-medium similarity renormalization group~(VS-IMSRG) calculations employing different interactions, provide direct constraints on the $^{81}$Zn wave function. These results point to the necessity of cross-shell excitations across the $^{78}$Ni core and show that, even in the immediate vicinity of doubly magic $^{78}$Ni, the competition between single-particle and collective degrees of freedom remains essential.

\begin{figure}[t!]
\includegraphics[width=0.49\textwidth]{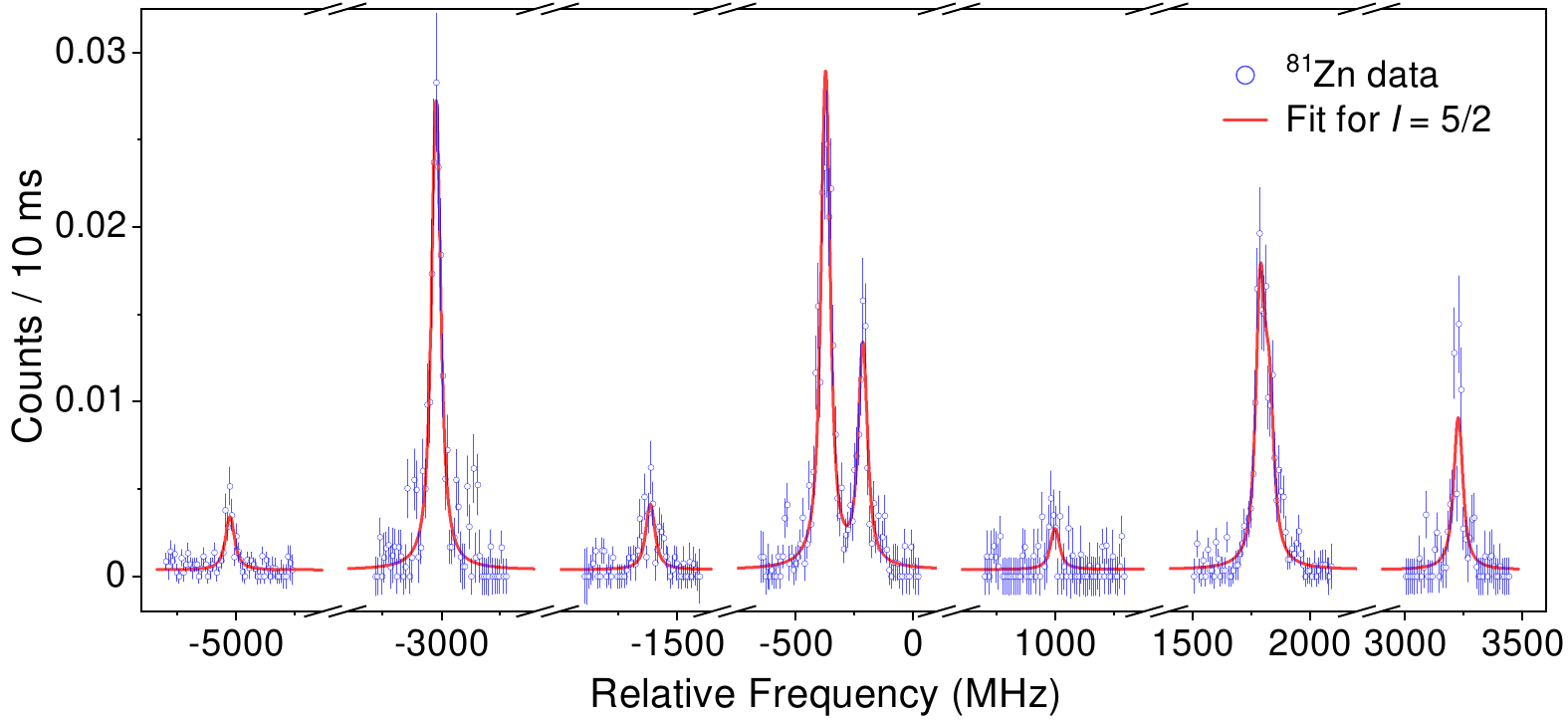}
\vspace{-7mm}
\caption{\label{fig:fig1}\footnotesize{Hyperfine structure spectrum of $^{81}$Zn measured at CRIS-ISOLDE for the $4s4p$ $^3P_2$ $\to$ $4s5s$ $^3S_1$ transition. The data are fitted with hyperfine-structure Hamiltonians using Voigt profiles for each hyperfine component (red line).}}
\vspace{-3mm}
\end{figure}

\begin{table*}[!htb]
\centering
\begin{threeparttable}
\caption{\label{tab:table1}\footnotesize{The hyperfine structure constants $A$, $B$ of the ${^3\!}P_{2}$ and ${^3\!}S_{1}$ atomic states for $^{81}$Zn and $^{75\rm{g,m}}$Zn, and the extracted magnetic and quadrupole moments using the high-precision literature values of $^{67}$Zn~\cite{67Zn-HFS-1962,67Zn-M1-1967} as references.}}
\renewcommand*{\arraystretch}{1.2}
\setlength{\tabcolsep}{3.5mm}{
\begin{tabular}{cc| lll|lll}
\hline\hline
$A$ & $I^{\pi}$ &$A({^3\!}P_{2} )$ (MHz)& $A({^3\!}S_{1})$ (MHz)&$B({^3\!}P_{2})$ (MHz)&$\mu$ ($\mu_{\rm N}$)& $Q_{\rm s}$(b) & Note \\
\hline
67  & 5/2$^{-}$& $+$531.987(5)~\tnote{a}& & $+$35.806(5)~\tnote{a} & $+$0.875479(9)~\tnote{b}& $+$0.122(10)\tnote{c}& \multirow{3}{*}{Literature }\\\
75  & 7/2$^{+}$& $-$342.3(4)~\tnote{c}& $-$815.4(6)~\tnote{c}& $+$48(4)~\tnote{c}& $-$0.7887(9)~\tnote{c}& $+$0.16(2)~\tnote{c}& \\
75$^{m}$ &1/2$^{-}$ & $+$1696(3)~\tnote{c}& $+$4034(6)~\tnote{c}& & $+$0.5580(9)~\tnote{c} & & \\
\hline
75 & 7/2$^{+}$ & $-$342.0(2)& $-$814.9(3)& $+$43(2)& $-$0.7880(5) & $+$0.15(1)& \multirow{3}{*}{This work}	\\
75$^{m}$&1/2$^{-}$ & $+$1696(1)&$+$4043(3)& & $+$0.5582(3)& & \\
81 & 5/2$^{+}$& $-$576.4(4)& $-$1375.2(6)& $-$82(3)&$-$0.9486(7) & $-$0.28(3)& \\
\hline
\end{tabular}}
\begin{tablenotes}
\item[a]\footnotesize{Values for the $^{67}$Zn are taken from Ref.~\cite{67Zn-HFS-1962}.}
\item[b]\footnotesize{Value for the $^{67}$Zn is taken from Ref.~\cite{67Zn-M1-1967}.}
\item[c]\footnotesize{Values for the $^{67}$Zn and $^{75\rm{g,m}}$Zn are taken from Ref.~\cite{Zn-moment}.}
\end{tablenotes}
\vspace{-3mm}
\end{threeparttable}
\end{table*}

Neutron-rich zinc isotopes were produced at ISOLDE-CERN by impinging 1.4-GeV protons onto a thick UCx target. To suppress the near-stability isobar contamination, a proton to neutron converter~\cite{NeutronConvertorISOLDE} and a quartz transfer line~\cite{QuartzLineISOLDE} were employed. The produced $^{81}$Zn ($T_{1/2}$ = 299.4~ms) nuclei were selectively ionized using the resonance ionization laser ion source~\cite{RILIS_ISOLDE}, accelerated to 30~keV and subsequently separated by the high-resolution mass separator. The $^{81}$Zn ions were then cooled and trapped in a gas-filled radio frequency linear Paul trap~\cite{ISCOOL_ISOLDE}, delivering ion bunches with a temporal width less than 5 $\mu$s at a repetition rate of 100~Hz to the CRIS setup. The average $^{81}$Zn ion rate delivered to CRIS was below 1000 ions/s. Further details of the CRIS setup can be found in Refs.~\cite{CRIS_Setup2020,VoltageScan_CRIS}. The bunched $^{81}$Zn ion beams were first neutralized in flight in a charge-exchange cell (CEC) containing sodium vapor, with an estimated overall neutralization efficiency of about 50\%. Less than one third of the neutralized atoms are expected to populate the metastable $4s4p$ $^3$P$_2$ state~\cite{CEC}. Residual ions were removed downstream of the CEC with electric deflectors. The resulting atomic bunches were delivered to the laser-atom interaction region, maintained at an ultra-high vacuum of $\sim10^{-10}$ mbar to minimize non-resonant collisional ionization. There, the $^{81}$Zn atoms were spatially overlapped and temporally synchronized with three 100-Hz pulsed lasers. A frequency-doubled, narrow-band 481-nm laser was used to resonantly excite the $4s4p$ $^3$P$_2\rightarrow4s5s$ $^3$S$_1$ transition, followed by stepwise excitation and ionization using a 691-nm dye laser and a 1064-nm Nd:YAG laser, respectively. During the measurement, the frequency of the first-step narrow-band laser was stabilized using a wavelength meter, while the kinetic energy of the $^{81}$Zn ions was varied by applying a scanning voltage to the CEC~\cite{VoltageScan_CRIS}. The hyperfine-structure (hfs) spectrum was obtained by Doppler tuning~\cite{PPNP2023,VoltageScan_CRIS}, with the resonantly ionized $^{81}$Zn ions recorded as a function of the CEC tuning voltage, which was then converted to a frequency scale via the Doppler shift.

\begin{figure*}[t!]
\includegraphics[width=0.99\textwidth]{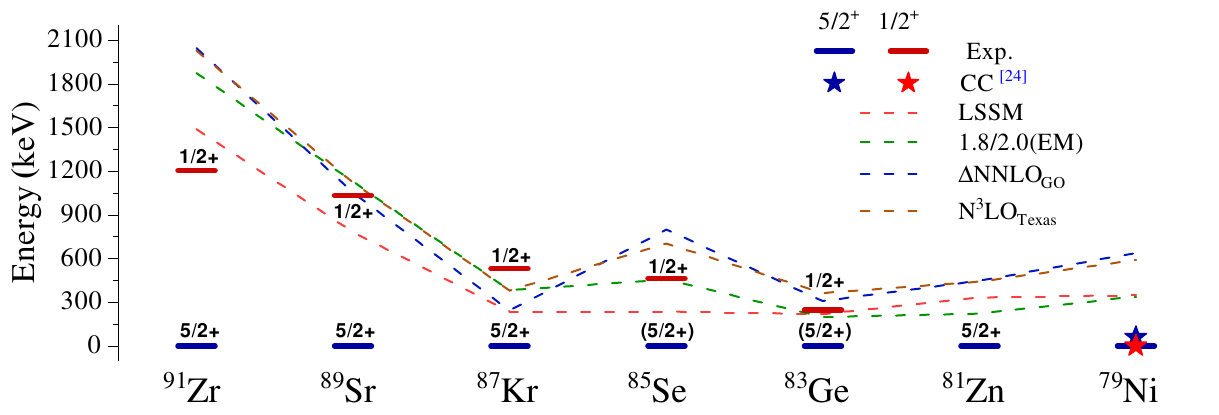}
\vspace{-3mm}
\caption{\footnotesize{Experimental energies of the low-lying 1/2$^+$ and 5/2$^+$ states (solid line) of $N = 51$ isotones between $Z=28$ and $Z=40$, compared to LSSM and {\it ab initio} VS-IMSRG calculations (dashed line) using three chiral interactions 1.8/2.0(EM), $\Delta$NNLO$_{\rm GO}$, and N$^3$LO$_{\rm Texas}$. A $^{52}$Ca core was used in all calculations, since excitations across the $^{78}$Ni core are required to reproduce the observed nuclear moments, as discussed below. The blue solid line in $^{81}$Zn represents the state with $5/2^{+}$ spin unambiguously confirmed in this work. The stars mark the $1/2^{+}$ ground state and $5/2^{+}$ excited state in $^{79}$Ni predicted by {\it ab initio} CC calculations~\cite{Gaute-78Ni}.}}
\label{fig:levels}
\vspace{-3mm}
\end{figure*}

The resulting hfs spectrum shown in Fig.~\ref{fig:fig1} was fitted with Voigt profiles using the $\chi^2$ minimization approach implemented in SATLAS2~\cite{SATLAS2}. For the $4s4p$ $^3P_2$ $\to$ $4s5s$ $^3S_1$ transition employed here, a nuclear spin $I = 1/2$ would give rise to only three allowed hyperfine transitions, whereas $I \geq 5/2$ yields nine. Assignments of $I=5/2,~7/2$ and 9/2 were therefore tested against the measured $^{81}$Zn spectrum. Only the $I=5/2$ assignment reproduces all observed resonance peaks~(Fig.~\ref{fig:fig1}) with a satisfactory reduced $\chi^2\approx1.0$, while the other spin assignments fail to reproduce the full spectral pattern. Furthermore, under the $I=5/2$ assignment, the extracted ratio of the magnetic hyperfine constants, $A(^3S_1)/A(^3P_2) = 2.386(2)$, agrees with the average ratio, $A(^3S_1)/A(^3P_2) = 2.382(2)$, measured for other zinc isotopes, for which the hyperfine anomaly is expected to be negligible~\cite{73Zn-moment}, providing an independent validation of the $I=5/2$ spin assignment. The ground- state spin of $^{81}$Zn is thus unambiguously determined to be $5/2$. Combined with the positive parity expected from systematics of the $N=51$ isotones and supported by the measured negative sign of the $^{81}$Zn magnetic moment discussed below, this establishes $I^\pi=5/2^+$. A broad frequency scan was also performed to search for the anticipated $I = 1/2$ state. The search region was defined based on the expected hyperfine pattern for $I=1/2$, the predicted magnetic moment, and a conservative range of possible isomer shifts. No corresponding resonance structure was observed over this extended frequency range. This suggests that the $1/2^+$ state in $^{81}$Zn has either a significantly shorter lifetime, a substantially lower production yield at ISOLDE, or a combination of both, placing it beyond the sensitivity of the present CRIS measurement.

The extracted hfs constants $A(^3P_2)$, $A(^3S_1)$, and $B(^3P_2)$ for $^{81}$Zn are summarized in Table~\ref{tab:table1}. As a benchmark, the hfs spectra of $^{75\rm{g,m}}$Zn were also measured, yielding hfs constants in good agreement with previous collinear laser-spectroscopy results~\cite{Zn-moment}, as also listed in Table~\ref{tab:table1}. The magnetic moment $\mu$ and spectroscopic quadrupole moment $Q_s$ of $^{81}$Zn were then determined from \mbox{$\mu= \mu_{\rm{ref}}IA/(I_{\rm{ref}}A_{\rm{ref}}$)} and \mbox{$Q_s= Q_{\rm s,ref}B/B_{\rm{ref}}$}, respectively, using the literature hfs constants and electromagnetic moments of $^{67}$Zn~\cite{67Zn-HFS-1962,67Zn-M1-1967} as the reference. The resulting moments are summarized in Table~\ref{tab:table1}. 

Our measurement unambiguously establishes $I^\pi=5/2^+$ for the $^{81}$Zn ground state, settling an assignment that had remained unsolved after nearly two decades of $\beta$-decay spectroscopy~\cite{beta-decay-1,beta-decay-2,beta-decay-3}. Figure~\ref{fig:levels} (solid lines) shows the experimental energy systematics of the low-lying 1/2$^+$ and 5/2$^+$ states in the $N=51$ isotones from the $Z=40$ sub-shell towards the $Z=28$ shell closure. The persistence of the $5/2^+$ ground state at $Z=30$ suggests that the predicted $\nu 2d_{5/2}$--$\nu 3s_{1/2}$ inversion has not yet occurred in $^{81}$Zn~\cite{SM-51N,Gaute-78Ni}. The present result therefore establishes $^{81}$Zn as a key experimental benchmark for the evolution of these neutron orbitals along the $N=51$ isotonic chain. 

\begin{figure*}[htb!]
\includegraphics[width=0.95\textwidth]{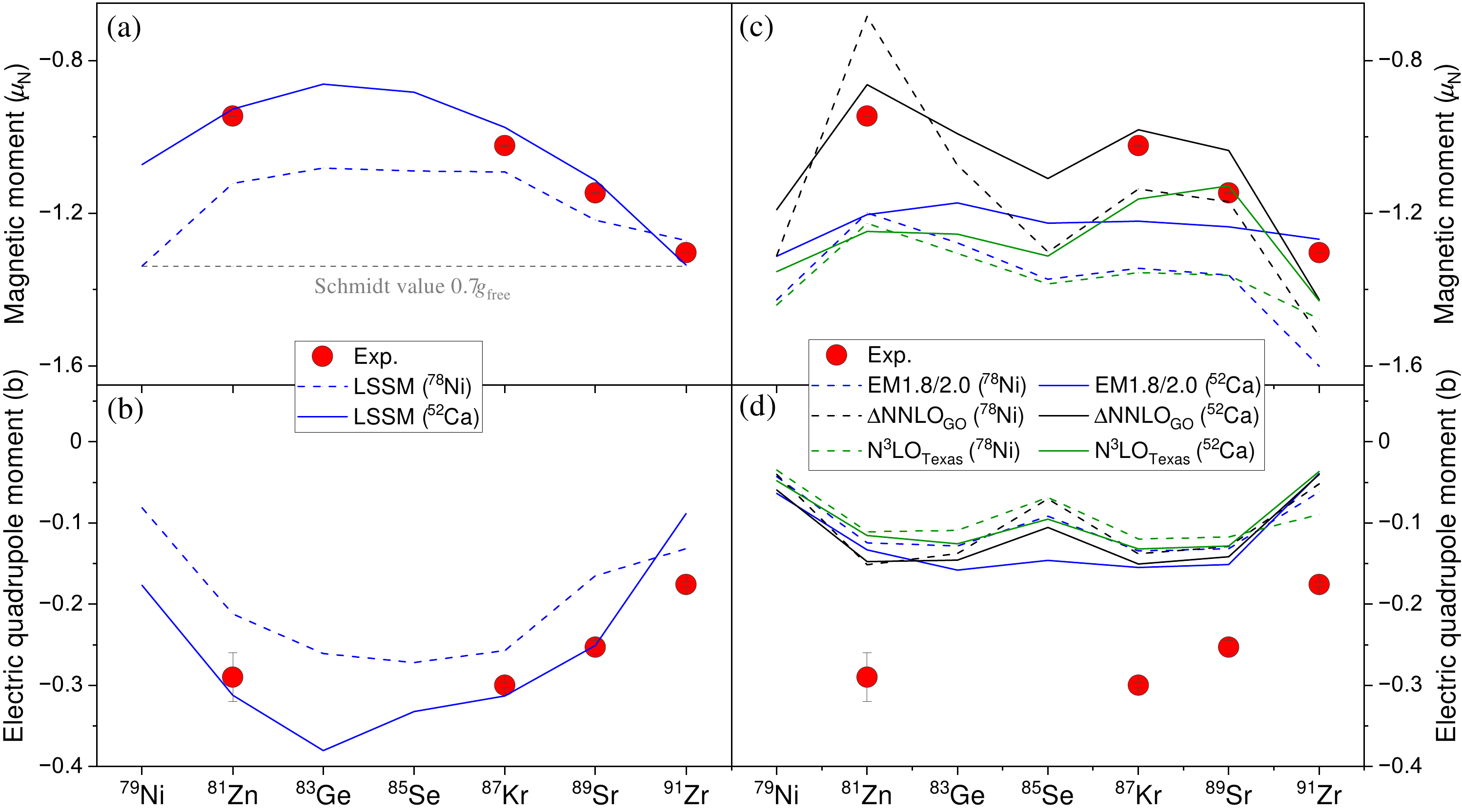}
\vspace{-2mm}
\caption{\footnotesize{(a) Magnetic and (b) quadrupole moments of the 5/2$^+$ ground states of $N = 51$ isotones compared to the LSSM calculations using $^{78}$Ni and $^{52}$Ca core. The effective $g$-factor ($g_{\rm{eff}}=0.7g_{\rm{free}}$) and effective charges ($e_\pi=1.5e$, $e_\nu=0.5e$) are used in LSSM calculations. (c) Magnetic and (d) quadrupole moments of 5/2$^+$ ground state of $N = 51$ isotones compared to {\it ab initio} VS-IMSRG calculations based on 1.8/2.0(EM), $\Delta$NNLO$_{\rm GO}$ and N$^3$LO$_{\rm Texas}$ chiral interactions using the $^{78}$Ni and $^{52}$Ca core. Two-body currents are taken into account~\cite{PhysRevLett.132.232503}, and free $g$-factors and bare nucleon charges are used in VS-IMSRG calculations. The experimental moments of $^{87}$Kr, $^{89}$Sr, $^{91}$Zr are taken from Refs.~\cite{Kr-moments-NPA-1995,Sr-moments-PRC-1990,Zr-moments-PR-1957,Zr-Qmoment-CPL-2000}.}}
\label{fig:moments}
\vspace{-3mm}
\end{figure*}

With the magnetic dipole and electric quadrupole moments of $^{81}$Zn determined for the first time, the experimental systematics of electromagnetic moments along the $N=51$ isotonic chain can now be extended to $Z=30$, providing a sensitive probe of the evolution of the nuclear moments toward $^{78}$Ni. As shown in Fig.~\ref{fig:moments}(a), the magnetic moments of the $5/2^+$ states near the $Z=40$ subshell lie close to the effective Schmidt value for a $\nu 2d_{5/2}$ neutron, consistent with a predominantly single-particle configuration. With decreasing $Z$, the moments progressively deviate from this value, with a particularly pronounced deviation in $^{81}$Zn despite its proximity to the $Z=28$ shell closure, indicating increasing configuration mixing and correlations beyond the single-particle picture. The quadrupole moments shown in Fig.~\ref{fig:moments}(b) provide complementary evidence. The $Q_s$ is relatively small near the $Z=40$ subshell and increases in magnitude with decreasing $Z$, signaling an enhanced contribution of collective configurations. Remarkably, the $Q_s$ value for $^{81}$Zn is, within uncertainties, comparable to that of $^{87}$Kr in the more open proton shell. Thus, despite only two protons above $Z=28$ and one neutron beyond $N=50$, $^{81}$Zn exhibits substantial configuration mixing and collective correlations.

To elucidate the microscopic origin of these correlations revealed by the electromagnetic moments, we performed LSSM and \textit{ab initio} VS-IMSRG calculations using two model spaces with $^{52}$Ca and $^{78}$Ni as cores. Details of these calculations are presented in the \textbf{End Matter}. As shown in Fig.~\ref{fig:moments}(a), the LSSM calculations based on the $^{52}$Ca core systematically reproduce the experimental $\mu$ across the $N=51$ isotones and, most notably, the newly measured value for $^{81}$Zn. In contrast, the restricted $^{78}$Ni-core calculation shows a pronounced discrepancy for $^{81}$Zn, demonstrating the increasing importance of nucleon excitations across the nominal $^{78}$Ni core. This interpretation is further supported by the calculated cross-shell configuration weights involving excitations across the $Z = 28$ and/or $N =50$ shell closures in the enlarged $^{52}$Ca-core model space (see \textbf{End Matter}), which amount to about 34.8\% for the 5/2$^+$ ground state of $^{81}$Zn. The quadrupole moments shown in Fig.~\ref{fig:moments}(b) provide complementary evidence for the importance of these cross-shell correlations. The enlarged-model space calculation reproduces the enhanced $|Q_s|$ of $^{81}$Zn and the overall experimental systematics, whereas the restricted $^{78}$Ni-core calculation underestimates the observed quadrupole collectivity. The exception at $^{91}$Zr likely reflects limitations of the adopted proton $pf$ valence space toward larger $Z$. Taken together, the simultaneous description of $\mu$ and $Q_s$ and the sizable calculated cross-shell contribution favor an $^{81}$Zn ground state wave function with substantial configuration mixing beyond a simple picture of a $\nu 2d_{5/2}$ neutron coupled to an inert $^{78}$Ni core. These results highlight the importance of correlations across the $Z=28$ and $N=50$ shell closures even in the immediate vicinity of $^{78}$Ni, in line with previous studies of the shell evolution in this region~\cite{Taniuchi2019-Ni,deGroote2020-Cu,deGroote2017-PRC-Cu,Yang2016-Zn79}.

The VS-IMSRG calculations shown in Fig.~\ref{fig:moments}~(c) also exhibit a pronounced model-space dependence for the magnetic moments. For all three chiral interactions 1.8/2.0(EM)~\cite{PhysRevC.83.031301,PhysRevC.96.014303}, $\Delta$NNLO$_{\rm GO}$~\cite{PhysRevC.102.054301} and N$^3$LO$_{\rm Texas}$~\cite{hu2026-texas} used here, calculations employing the enlarged $^{52}$Ca-core valence space provide an improved description of the experimental magnetic moments compared with the corresponding $^{78}$Ni-core calculations, further supporting the importance of excitations across the $Z=28$ and $N=50$ shell closures. This improvement is particularly evident for $^{81}$Zn with the $\Delta$NNLO$_{\rm GO}$ interaction, for which the calculation using the $^{52}$Ca core shifts toward the experimental value. In contrast, the calculated quadrupole moments shown in Fig.~\ref{fig:moments}~(d) depend only weakly on the interactions and adopted valence spaces, with all calculations substantially underestimating the experimental $|Q_s|$. A similar underestimation of quadrupole collectivity has already been observed in previous VS-IMSRG studies in other mass regions~\cite{Sc-moment,In-moment} and has been attributed to missing higher-order collective correlations associated with many-particle--many-hole excitations that are not fully captured within the truncation at the two-body level, IMSRG(2)~\cite{PhysRevC.105.034333}. The discrepancy in $Q_s$ therefore points to the need for further developments in VS-IMSRG treatments of collective $E2$ correlations.

Having established the importance of cross-shell correlations from the electromagnetic moments, we now return to the evolution of the low-lying $5/2^+$ and $1/2^+$ states along the $N=51$ isotonic chain. The calculations shown in Fig.~\ref{fig:levels} therefore employ only the enlarged valence space above a $^{52}$Ca core, allowing excitations across both the $Z=28$ and $N=50$ shell closures. Both the LSSM and VS-IMSRG calculations reproduce the overall experimental evolution of the two states, including the rapid lowering of the $1/2^+$ state toward $Z=28$. Nevertheless, all calculations yield a $5/2^+$ ground state for both $^{81}$Zn and $^{79}$Ni, consistent with the $^{81}$Zn spin established in this work, and show no $\nu 2d_{5/2}$--$\nu 3s_{1/2}$ inversion at $Z=28$. This contrasts with earier SM calculations using the effective interaction jj45pna and CC calculations that predicted a $1/2^+$ ground state in $^{81}$Zn and/or $^{79}$Ni~\cite{SM-51N,Gaute-78Ni}. These results reveal a subtle picture of shell evolution near $^{78}$Ni. While the $\nu2d_{5/2}$--$\nu3s_{1/2}$ level ordering remains robust at $Z=30$, the $^{81}$Zn ground state moments already contain substantial correlations beyond an inert-$^{78}$Ni-core picture. The preservation of this ordering therefore does not imply a simple single-particle structure, but instead highlights the interplay between shell evolution and cross-shell many-body correlations. Whether the inversion occurs at $Z=28$ remains experimentally unsolved, making a direct determination of the $^{79}$Ni ground state spin a decisive future test. It is also worth noting that the VS-IMSRG calculations exhibit a common kink around $Z=34$ ($^{85}$Se) in the magnetic moments, quadrupole moments, and excitation energies (Figs.~\ref{fig:levels} and~\ref{fig:moments}). The consistency of this feature across different observables may indicate a weak subshell effect at $Z=34$, which calls for further theoretical and experimental investigation around $^{85}$Se.

In summary, the ground state properties of $^{81}$Zn have been directly determined using collinear resonance ionization spectroscopy at CERN-ISOLDE. Its spin and parity are unambiguously established as $I^\pi=5/2^+$, demonstrating that the predicted $\nu2d_{5/2}$--$\nu3s_{1/2}$ inversion has not yet occurred at $Z=30$. The measurements of its magnetic dipole and electric quadrupole moments reveal pronounced deviations from a simple single-particle picture, indicating substantial configuration mixing and collective correlations in the $^{81}$Zn ground state wave function. Comparisons with LSSM and \textit{ab initio} VS-IMSRG calculations show that descriptions allowing excitations across the $Z=28$ and $N=50$ shell closures provide an improved account of the experimental moments. These results therefore demonstrate the importance of cross-shell correlations in the vicinity of doubly magic $^{78}$Ni. At the same time, the persistence of the $\nu2d_{5/2}$--$\nu3s_{1/2}$ level ordering at $Z=30$ shows that a robust single-particle ordering can coexist with substantial correlations, highlighting the interplay between shell evolution and collective degrees of freedom near doubly magic $^{78}$Ni. A direct determination of the $^{79}$Ni ground state spin in the future will provide a decisive experimental test of the level ordering at $Z=28$.

\vspace{2mm}

We acknowledge the support of the ISOLDE collaboration and technical teams. X.F.Y. acknowledges support from the National Key R\&D Program of China (Contract No:2023YFA1606403, 2022YFA1605100), the National Natural Science Foundation of China (Contract No: 12305122); G.N. and T.E.C. acknowledge the BriX Research Program No. P7/12; G.N., R.d.G., A.K., T.E.C. acknowledge support from KU Leuven project C14/22/104 and from the FWO Vlaanderen IRI project I001323N, as well as projects  G053221N, G080022N, 1167322 and G0G3121N; K.F. acknowledges support from the UK Science and Technology Facilities Council grants ST/P004423/1, ST/V001116/1 and ST/Y000323/1; K.M.L.acknowledges support from the Royal Society Grant No. DHF/R1/231007 and UKRI under the Horizon Europe funding Guarantee Grant No. EP/Y036816/1 (ESPEN); R.F.G.R. acknowledges support from the Office of Science U.S. Department of Energy, under the grant DE-SC0021176; R.L acknowledges support from the Romanian National Authority for Research through the Nucleu Project No. PN 23 21 01 02 and the Institute of Atomic Physics through the CERN/ISOLDE grant; J.D.H. acknowledges support from NSERC under grant SAPIN-2024-0003 and the Arthur B. McDonald Canadian Astroparticle Physics Research Institute; B.S.H acknowledges support from the National Natural Science Foundation of China under Grants No.12675150; the Fundamental Research Funds for the Central Universities, Peking University. ISOLDE experiment was supported by the EU Horizon2020 research and innovation programme through ENSAR2 654002; Computations were performed with an allocation of computing resources on Fir from The Digital Research Alliance of Canada and the High-Performance Computing Platform of Peking University;The IMSRG code used is \texttt{ragnar\_imsrg}~\cite{code-imsrg}. We appreciate Dr. Mikael Reponen from the University of Jyv\"askyl\"a regarding the laser system.

\bibliography{81Zn, Ref_Baishan}

@STRING {N = "Nat"}

@STRING {NP = "Nat. Phys"}

@STRING {NIMB = "Nucl. Instrum. Meth. B"}

@STRING {PLB = "Phys. Lett. B"}

@STRING {PPNP = "Prog. Part. Nucl. Phys"}

@STRING {PRC = "Phys. Rev. C"}

@STRING {PRL = "Phys. Rev. Lett"}

@STRING {EPJA = "Eur. Phys. J. A"}

@STRING {PR = "Phys. Rev"}

@STRING {PLA = "Phys. Lett. A"}

@STRING {RMP = "Rev. Mod. Phys"}

@article{In-moment,
  title = {Electromagnetic properties of indium
isotopes illuminate the doubly magic
character of $^{100}\mathrm{Sn}$},
  author = {Karthein, J. and Ricketts, C.M. and Garcia Ruiz, R.F. and others},
  journal = {Nat. Phys.},
  volume = {20},
  pages = {1719-1725},
  year = {2024},
  doi = {10.1038/s41567-024-02612-y},
  url = {https://doi.org/10.1038/s41567-024-02612-y}
}

@article{Sc-moment,
title = {Electromagnetic moments of scandium isotopes and ${N}=28$ isotones in the distinctive 0f7/2 orbit},
journal = {Physics Letters B},
volume = {829},
pages = {137064},
year = {2022},
issn = {0370-2693},
doi = {https://doi.org/10.1016/j.physletb.2022.137064},
url = {https://www.sciencedirect.com/science/article/pii/S0370269322001988},
author = {S.W. Bai and \'A. {Koszor\'us} and B.S. Hu and others},
}

@article{beta-decay-1,
  title = {Low-energy states of $^{81}\mathrm{Ga}$: Proton structure of the nuclei close to $^{78}\mathrm{Ni}$},
  author = {Verney, D. and Ibrahim, F. and Bourgeois, C. and others},
  collaboration = {PARRNe Collaboration},
  journal = {Phys. Rev. C},
  volume = {76},
  issue = {5},
  pages = {054312},
  numpages = {7},
  year = {2007},
  month = {Nov},
  publisher = {American Physical Society},
  doi = {10.1103/PhysRevC.76.054312},
  url = {https://link.aps.org/doi/10.1103/PhysRevC.76.054312}
}

@article{beta-decay-2,
  title = {$\ensuremath{\beta}$ decay of $^{81}\mathrm{Zn}$ and migrations of states observed near the $N=50$ closed shell},
  author = {Padgett, S. and Madurga, M. and Grzywacz, R. and others},
  journal = {Phys. Rev. C},
  volume = {82},
  issue = {6},
  pages = {064314},
  numpages = {9},
  year = {2010},
  month = {Dec},
  publisher = {American Physical Society},
  doi = {10.1103/PhysRevC.82.064314},
  url = {https://link.aps.org/doi/10.1103/PhysRevC.82.064314}
}

@article{beta-decay-3,
  title = {Fast-timing study of $^{81}\mathrm{Ga}$ from the $\ensuremath{\beta}$ decay of $^{81}\mathrm{Zn}$},
  author = {Paziy, V. and Fraile, L. M. and Mach, H. and other},
  journal = {Phys. Rev. C},
  volume = {102},
  issue = {1},
  pages = {014329},
  numpages = {18},
  year = {2020},
  month = {Jul},
  publisher = {American Physical Society},
  doi = {10.1103/PhysRevC.102.014329},
  url = {https://link.aps.org/doi/10.1103/PhysRevC.102.014329}
}

@article{PhysofExoticNuclei,
	title = {Physics of exotic nuclei},
	volume = {7},
	issn = {2522-5820},
	url = {https://doi.org/10.1038/s42254-024-00782-5},
	doi = {10.1038/s42254-024-00782-5},
	journal = {Nature Reviews Physics},
	author = {Ye, Yan Lin and Yang, Xiao Fei and Sakurai, Hiroyoshi and Hu, Bai Shan},
	month = jan,
	year = {2025},
	pages = {21--37},
}

@article{ShellEvolution-Ostuka,
    author = {Otsuka, T. and Gade, A. and Sorlin, O. and others},
    title = {Evolution of shell structure in exotic nuclei},
    journal = RMP,
    volume = {92},
    pages = {015002},
    year = {2020},
    doi = {10.1103/RevModPhys.92.015002},
    url = {https://link.aps.org/doi/10.1103/RevModPhys.92.015002}
}

@article{NewMagicNumber-Sorlin,
title = {Nuclear magic numbers: New features far from stability},
journal = {Progress in Particle and Nuclear Physics},
volume = {61},
number = {2},
pages = {602-673},
year = {2008},
issn = {0146-6410},
doi = {https://doi.org/10.1016/j.ppnp.2008.05.001},
url = {https://www.sciencedirect.com/science/article/pii/S0146641008000380},
author = {O. Sorlin and M.-G. Porquet},
}

@article{WaitingPoint-2,
    author = {Hosmer, P.T. and Schatz, H. and Aprahamian, A. and others},
    title = {Half-Life of the Doubly Magic $r$-Process Nucleus $^{78}\mathrm{Ni}$},
    journal = PRL,
    year = {2005},
    volume = {94},
    pages = {112501},
    doi = {10.1103/PhysRevLett.94.112501},
    url = { https://doi.org/10.1103/PhysRevLett.94.112501}
}

@article{Nies2023-Zn79m,
  title = {Further Evidence for Shape Coexistence in ${^{79}\mathrm{Zn}}^{m}$ near Doubly Magic $^{78}\mathrm{Ni}$},
  author = {Nies, L. and Canete, L. and Dao, D. D. and others},
  journal = {Phys. Rev. Lett.},
  volume = {131},
  issue = {22},
  pages = {222503},
  numpages = {7},
  year = {2023},
  month = {Nov},
  publisher = {American Physical Society},
  doi = {10.1103/PhysRevLett.131.222503},
  url = {https://link.aps.org/doi/10.1103/PhysRevLett.131.222503}
}

@Article{Flanagan2009-Cu,
  Title = "Nuclear spins and magnetic moments of {$^{71,73,75}$Cu}: inversion of the $\pi2p_{3/2}$ and $\pi1f_{5/2}$ levels in {$^{75}$Cu}",
    Author = "Flanagan, K. T. and Vingerhoets, P. and Avgoulea, M. and others",
  Journal = PRL,
  Year = "2009",
  Volume = "103",
  Pages = "142501",
  doi = "10.1103/PhysRevLett.103.142501",
  url = "https://link.aps.org/doi/10.1103/PhysRevLett.103.142501"
}

@Article{Cheal2010-Ga81,
  Title = "Nuclear spins and moments of {Ga} isotopes reveal sudden structural changes between {$N=40$ and $N=50$}",
  Author = "Cheal, B. and Man\'e, E. and Billowes, J. and others",
  Journal = PRL,
  Year = "2010",
  Volume = "104",
  Pages = "252502",
  doi = "10.1103/PhysRevLett.104.252502",
  url = "https://link.aps.org/doi/10.1103/PhysRevLett.104.252502"
}

@Article{deGroote2017-PRC-Cu,
  Title = "Dipole and quadrupole moments of {$^{73-78}$Cu} as a test of the robustness of the {$Z=28$} shell closure near {$^{78}$Ni}",
  Author = "de Groote, R. P. and Billowes, J. and Binnersley, C. L. and others",
  Journal = PRC,
  Year = "2017",
  Volume = "96",
  Pages	= "041302",
  Doi = "10.1103/PhysRevC.96.041302",
  Url = "https://link.aps.org/doi/10.1103/PhysRevC.96.041302"
}

@Article{deGroote2020-Cu,
  Title = "Measurement and microscopic description of odd-even staggering of charge radii of exotic copper isotopes",
  Author = "de Groote, R. P. and Billowes, J. and Binnersley, C. L. and others",
  Journal = NP,
  Year = "2020",
  Volume = "16",
  Pages	= "620-624",
  doi = "10.1038/s41567-020-0868-y",
  url = "https://doi.org/10.1038/s41567-020-0868-y"
}

@article{Yang2016-Zn79,
  title = {Isomer Shift and Magnetic Moment of the Long-Lived $1/{2}^{+}$ Isomer in $_{30}^{79}\mathrm{Zn}_{49}$: Signature of Shape Coexistence near $^{78}\mathrm{Ni}$},
  author = {Yang, X. F. and Wraith, C. and Xie, L. and others},
  journal = PRL,
  volume = {116},
  issue = {18},
  pages = {182502},
  numpages = {6},
  year = {2016},
  month = {May},
  publisher = {American Physical Society},
  doi = {10.1103/PhysRevLett.116.182502},
}

@article{PhysRevC.111.044308,
  title = {Realistic shell model calculations for shell evolution near $^{78}\mathrm{Ni}$},
  author = {Wang, X. P. and Li, J. G. and Chen, N.},
  journal = {Phys. Rev. C},
  volume = {111},
  issue = {4},
  pages = {044308},
  numpages = {7},
  year = {2025},
  month = {Apr},
  publisher = {American Physical Society},
  doi = {10.1103/PhysRevC.111.044308},
  url = {https://link.aps.org/doi/10.1103/PhysRevC.111.044308}
}

@article{Olivier2017-Cu79,
  title = {Persistence of the ${Z}=28$ Shell Gap Around $^{78}\mathrm{Ni}$: First Spectroscopy of $^{79}\mathrm{Cu}$},
  author = {Olivier, L. and Franchoo, S. and Niikura, M. and },
  journal = {Phys. Rev. Lett.},
  volume = {119},
  issue = {19},
  pages = {192501},
  numpages = {6},
  year = {2017},
  month = {Nov},
  publisher = {American Physical Society},
  doi = {10.1103/PhysRevLett.119.192501},
  url = {https://link.aps.org/doi/10.1103/PhysRevLett.119.192501}
}

@Article{Welker2017-Cu,
  Title = {Binding energy of {$^{79}$Cu}: probing the structure of the doubly magic {$^{78}$Ni} from only one proton away},
  Author = {Welker, A. and Althubiti, N. A. S. and Atanasov, D. and others},
  Journal = PRL,
  Year = {2017},
  Volume = {119},
  Pages = {192502},
  doi = {10.1103/PhysRevLett.119.192502},
  url = {https://link.aps.org/doi/10.1103/PhysRevLett.119.192502}
}

@Article{Taniuchi2019-Ni,
  Title = "{$^{78}$Ni} revealed as a doubly magic stronghold against nuclear deformation",
  Author = "Taniuchi, R. and Santamaria, C. and Doornenbal, P. and others",
  Journal = N,
  Year = "2019",
  Volume = "569",
  Pages = "53-58",
  doi = "10.1038/s41586-019-1155-x",
  url = "https://doi.org/10.1038/s41586-019-1155-x"
}

@article{PPNP2023,
title = {Laser spectroscopy for the study of exotic nuclei},
journal = PPNP,
volume = {129},
pages = {104005},
year = {2023},
issn = {0146-6410},
author = {X.F. Yang and S.J. Wang and S.G. Wilkins and others},
doi = {https://doi.org/10.1016/j.ppnp.2022.104005},
url = {https://www.sciencedirect.com/science/article/pii/S0146641022000631}
}

@article{SATLAS2,
title = {{SATLAS2}: An update to the package for analysis of counting data},
journal = {Comput. Phys. Commun},
volume = {297},
pages = {109053},
year = {2024},
issn = {0010-4655},
doi = {https://doi.org/10.1016/j.cpc.2023.109053},
url = {https://www.sciencedirect.com/science/article/pii/S0010465523003983},
author = {W. Gins and B. {van den Borne} and R.P. {de Groote} and others},
}

@ARTICLE{VoltageScan_CRIS,
   Title = "Voltage scanning and technical upgrades at the Collinear Resonance Ionization Spectroscopy experiment",
   Author   = "Athanasakis-Kaklamanakis, M. and Reilly, J. R. and Koszor\'us, \'A. and others",
   Year     = "2023",
   Journal  = NIMB,
   Volume   = "541",
   Pages    = "86-89",
   doi = {10.1016/j.nimb.2023.04.054},
   url = {https://www.sciencedirect.com/science/article/pii/S0168583X2300188X}
}

@ARTICLE{CRIS_Setup2020,
   Title = "Optimising the Collinear Resonance Ionisation Spectroscopy ({CRIS}) experiment at CERN-ISOLDE",
   Author   = "A. R. {Vernon} and R. P. {de Groote} and J. {Billowes} and others",
   Year     = "2020",
   Journal  = NIMB,
   Volume   = "463",
   Pages    = "384-389",
   doi = {10.1016/j.nimb.2019.04.049},
   url = {https://www.sciencedirect.com/science/article/pii/S0168583X19302356}
}

@Article{Kaufmann2020-Ni,
  Title = "Charge radius of the short-lived {$^{68}$Ni} and correlation with the dipole polarizability",
  Author = "Kaufmann, S. and Simonis, J. and Bacca, S. and others",
  Journal = PRL,
  Year = "2020",
  Volume = "124",
  Pages = "132502",
  Doi = {10.1103/PhysRevLett.124.132502},
  Url = {https://link.aps.org/doi/10.1103/PhysRevLett.124.132502}
}

@article{Ni-radii,
  title = {Nuclear Charge Radii of the Nickel Isotopes $^{58\ensuremath{-}68,70}\mathrm{Ni}$},
  author = {Malbrunot-Ettenauer, S. and Kaufmann, S. and Bacca, S. and others},
  journal = {Phys. Rev. Lett.},
  volume = {128},
  issue = {2},
  pages = {022502},
  numpages = {8},
  year = {2022},
  month = {Jan},
  publisher = {American Physical Society},
  doi = {10.1103/PhysRevLett.128.022502},
  url = {https://link.aps.org/doi/10.1103/PhysRevLett.128.022502}
}

@article{Gaute-78Ni,
  title = {Structure of $^{78}\mathrm{Ni}$ from First-Principles Computations},
  author = {Hagen, G. and Jansen, G. R. and Papenbrock, T.},
  journal = PRL,
  volume = {117},
  issue = {17},
  pages = {172501},
  numpages = {7},
  year = {2016},
  month = {Oct},
  publisher = {American Physical Society},
  doi = {10.1103/PhysRevLett.117.172501},
  url = {https://link.aps.org/doi/10.1103/PhysRevLett.117.172501}
}

@article{SM-51N,
  title = {Shell Model Description of {$N = 51$} Isotones},
  author = {K. {Maurya} and P. C. {Srivastava} and I. {Mehrotra}},
  journal = {J. Appl. Phys.},
  volume = {3},
  issue = {4},
  pages = {52},
  year = {2013},
  doi = {10.9790/4861-0345259 },
  url = {https://www.iosrjournals.org/iosr-jap/papers/Vol3-issue4/G0345259.pdf}
}

@article{SM78Ni,
  title = {Shape Coexistence in $^{78}\mathrm{Ni}$ as the Portal to the Fifth Island of Inversion},
  author = {Nowacki, F. and Poves, A. and Caurier, E. and others},
  journal = PRL,
  volume = {117},
  issue = {27},
  pages = {272501},
  numpages = {5},
  year = {2016},
  month = {Dec},
  publisher = {American Physical Society},
  doi = {10.1103/PhysRevLett.117.272501},
  url = {https://link.aps.org/doi/10.1103/PhysRevLett.117.272501}
}

@article{2009N51,
  title = {Shell model description of zirconium isotopes},
  author = {Sieja, K. and Nowacki, F. and Langanke, K. and others},
  journal = PRC,
  volume = {79},
  issue = {6},
  pages = {064310},
  numpages = {9},
  year = {2009},
  month = {Jun},
  publisher = {American Physical Society},
  doi = {10.1103/PhysRevC.79.064310},
  url = {https://link.aps.org/doi/10.1103/PhysRevC.79.064310}
}

@article{expN51,
  title = {First study of the level structure of the r-process nucleus $^{83}\mathrm{Ge}$},
  author = {Thomas, J. S. and Bardayan, D. W. and Blackmon, J. C. and others},
  journal = PRC,
  volume = {71},
  issue = {2},
  pages = {021302},
  numpages = {4},
  year = {2005},
  month = {Feb},
  publisher = {American Physical Society},
  doi = {10.1103/PhysRevC.71.021302},
  url = {https://link.aps.org/doi/10.1103/PhysRevC.71.021302}
}

@article{81Zngs,
  title = {$\ensuremath{\beta}$ decay of $^{81}\mathrm{Zn}$ and migrations of states observed near the ${N}=50$ closed shell},
  author = {Padgett, S. and Madurga, M. and Grzywacz, R. and others},
  journal = PRC,
  volume = {82},
  issue = {6},
  pages = {064314},
  numpages = {9},
  year = {2010},
  month = {Dec},
  publisher = {American Physical Society},
  doi = {10.1103/PhysRevC.82.064314},
  url = {https://link.aps.org/doi/10.1103/PhysRevC.82.064314}
}

@article{NeutronConvertorISOLDE,
  title = {Experimental tests of an advanced proton-to-neutron converter at {ISOLDE-CERN}},
  author = {Gottberg, A. and Mendonca, T. M. and Luis, R. and others},
  journal = NIMB,
  volume = {336},
  pages = {143-148},
  year = {2014},
  doi = {10.1016/j.nimb.2014.04.026},
  url = {https://www.sciencedirect.com/science/article/pii/S0168583X1400528X}
}

@article{QuartzLineISOLDE,
  title = {Beam purification by selective trapping in the transfer line of an {ISOL} target unit},
  author = {Boucquerel, E. and Catherall, R. and Eller, E. and others},
  journal = NIMB,
  volume = {266},
  pages = {4298-4302},
  year = {2008},
  doi = {10.1016/j.nimb.2008.05.060},
  url = {https://www.sciencedirect.com/science/article/pii/S0168583X08007246}
}

@article{RILIS_ISOLDE,
  title = {Laser ion beam production at {CERN-ISOLDE}: New features – More possibilities},
  author = {Rothe, S and Day Goodacre, T. and Fedorov, D. V. and others},
  journal = NIMB,
  volume = {376},
  pages = {91-96},
  year = {2016},
  doi = {10.1016/j.nimb.2016.02.024},
  url = {https://www.sciencedirect.com/science/article/pii/S0168583X1600152X}
}

@article{ISCOOL_ISOLDE,
  title = {An ion cooler-buncher for high-sensitivity collinear laser spectroscopy at {ISOLDE}},
  author = {Man\'e, E. and Billowes, J. and Blaum, K. and others},
  journal = EPJA,
  volume = {42},
  pages = {503-507},
  year = {2009},
  doi = {10.1140/epja/i2009-10828-0},
  url = {https://doi.org/10.1140/epja/i2009-10828-0}
}

@article{67Zn-HFS-1962,
  title = {Hyperfine Structure of the $^{3}{P}$ States of $^{67}${Zn} and $^{25}\mathrm{Mg}$},
  author = {Lurio, A},
  journal = PR,
  volume = {126},
  pages = {1768-1773},
  year = {1962},
  doi = {10.1103/PhysRev.126.1768},
  url = {https://link.aps.org/doi/10.1103/PhysRev.126.1768}
}

@article{67Zn-M1-1967,
    author = {Spence, P. W. and McDermott, M. N.},
    title = {Optical orientation of $^{67}${Zn}},
    journal = PLA,
    volume = {24},
    pages = {430-431},
    year = {1967},
    doi = {10.1016/0375-9601(67)90222-8},
    url = {https://www.sciencedirect.com/science/article/pii/0375960167902228}
}

@article{Cu-radii2020,
  title={Measurement and microscopic description of odd--even staggering of charge radii of exotic copper isotopes},
  author={de Groote, RP and Billowes, Jon and Binnersley, Corry L and Bissell, Mark L and Cocolios, Thomas Elias and Goodacre, T Day and Farooq-Smith, Gregory James and Fedorov, DV and Flanagan, Kieran T and Franchoo, Serge and others},
  journal=NP,
  volume = {16},
  pages={620-624},
  year={2020},
  doi ={doi.org/10.1038/s41567-020-0868-y},
  url = {https://doi.org/10.1038/s41567-020-0868-y},
}

@article{Zn-radii,
title = {Nuclear charge radii of $^{62\ensuremath{-}80}\mathrm{Zn}$ and their dependence on cross-shell proton excitations},
journal = {Physics Letters B},
volume = {797},
pages = {134805},
year = {2019},
issn = {0370-2693},
doi = {https://doi.org/10.1016/j.physletb.2019.134805},
url = {https://www.sciencedirect.com/science/article/pii/S037026931930509X},
author = {L. Xie and X.F. Yang and C. Wraith and others},
}

@article{Zn-moment,
title = {Evolution of nuclear structure in neutron-rich odd-Zn isotopes and isomers},
journal = {Physics Letters B},
volume = {771},
pages = {385-391},
year = {2017},
issn = {0370-2693},
doi = {https://doi.org/10.1016/j.physletb.2017.05.085},
url = {https://www.sciencedirect.com/science/article/pii/S0370269317304483},
author = {C. Wraith and X.F. Yang and L. Xie and other},
}

@article{73Zn-moment,
  title = {Investigating the large deformation of the $5/{2}^{+}$ isomeric state in $^{73}\mathrm{Zn}$: An indicator for triaxiality},
  author = {Yang, X. F. and Tsunoda, Y. and Babcock, C. and others},
  journal = {Phys. Rev. C},
  volume = {97},
  issue = {4},
  pages = {044324},
  numpages = {8},
  year = {2018},
  month = {Apr},
  publisher = {American Physical Society},
  doi = {10.1103/PhysRevC.97.044324},
  url = {https://link.aps.org/doi/10.1103/PhysRevC.97.044324}
}

@article{ShellModel-JGL-2023,
    author = {Li, J. G.},
    title = {Merging of the island of inversion at ${N} = 40$ and ${N} = 50$},
    journal = PLB,
    volume = {840},
    pages = {137893},
    year = {2023},
    doi = {10.1016/j.physletb.2023.137893},
    url = {https://www.sciencedirect.com/science/article/pii/S0370269323002277}
}

@article{Kr-moments-NPA-1995,
title = {Laser-spectroscopy measurements of $^{72–96}${Kr} spins, moments and charge radii},
journal = {Nucl. Phys. A},
volume = {586},
number = {2},
pages = {219-239},
year = {1995},
issn = {0375-9474},
doi = {10.1016/0375-9474(94)00786-M},
url = {https://www.sciencedirect.com/science/article/pii/037594749400786M},
author = {M. {Keim} and E. {Arnold} and W. {Borchers} and others}
}

@article{Sr-moments-PRC-1990,
  title = {Erratum: Systematics of nuclear ground state properties in $^{78\ensuremath{-}100}\mathrm{Sr}$ by laser spectroscopy},
  author = {F. {Buchinger} and E. B. {Ramsay} and E. {Arnold} and others},
  journal = {Phys. Rev. C},
  volume = {42},
  issue = {6},
  pages = {2754--2754},
  numpages = {0},
  year = {1990},
  month = {Dec},
  publisher = {American Physical Society},
  doi = {10.1103/PhysRevC.42.2754},
  url = {https://link.aps.org/doi/10.1103/PhysRevC.42.2754}
}

@article{Zr-moments-PR-1957,
  title = {Magnetic Moment and Spin of $_{40}\mathrm{Zr}_{51}^{91}$},
  author = {E. {Brun} and J. {Oeser} and H. H. {Staub}},
  journal = {Phys. Rev.},
  volume = {105},
  issue = {6},
  pages = {1929--1930},
  numpages = {0},
  year = {1957},
  month = {Mar},
  publisher = {American Physical Society},
  doi = {10.1103/PhysRev.105.1929},
  url = {https://link.aps.org/doi/10.1103/PhysRev.105.1929}
}

@article{Zr-Qmoment-CPL-2000,
title = {The nuclear quadrupole moment of $^{91}${Zr} from molecular data for {ZrO} and {ZrS}},
journal = {Chem. Phys. Lett.},
volume = {318},
number = {1},
pages = {222-231},
year = {2000},
issn = {0009-2614},
doi = {10.1016/S0009-2614(00)00031-2},
url = {https://www.sciencedirect.com/science/article/pii/S0009261400000312},
author = {V. {Kell\"o} and P. {Pyykk\"o} and A. J. {Sadlej} and others}
}

@article{CEC,
title = {Simulation of the relative atomic populations of elements {$1 \leq Z \leq 89$} following charge exchange tested with collinear resonance ionization spectroscopy of indium},
journal = {Spectrochimica Acta Part B: Atomic Spectroscopy},
volume = {153},
pages = {61-83},
year = {2019},
issn = {0584-8547},
doi = {https://doi.org/10.1016/j.sab.2019.02.001},
url = {https://www.sciencedirect.com/science/article/pii/S0584854718305597},
author = {A.R. Vernon and J. Billowes and C.L. Binnersley and other},
}

@article{Stro22E2,
    author = "Stroberg, S. R. and Henderson, J. and Hackman, G. and Ruotsalainen, P. and Hagen, G. and Holt, J. D.",
    title = "{Systematics of E2 strength in the sd shell with the valence-space in-medium similarity renormalization group}",
    doi = "10.1103/PhysRevC.105.034333",
    journal = "Phys. Rev. C",
    volume = "105",
    number = "3",
    pages = "034333",
    year = "2022"
}

@misc{code-imsrg,
  author       = {S. R. Stroberg},
  title        = {\texttt{imsrg++}},
  year         = {2026},
  howpublished = {\url{https://github.com/ragnarstroberg/imsrg}},
  url          = {https://github.com/ragnarstroberg/imsrg}
}

@article{hu2026-texas,
  title = {Neutron Drip Line in Calcium Isotopes from a Chiral Interaction},
  author = {Hu, B. S. and Ekstr{\"o}m, A. and Forss{\'e}n, C. and Hagen, G. and Jiang, W. G. and Miyagi, T. and Papenbrock, T.},
  journal = {Phys. Rev. Lett.},
  volume = {137},
  issue = {14},
  pages = {142502},
  numpages = {10},
  year = {2026},
  month = {Sep},
  publisher = {American Physical Society},
  doi = {10.1103/pz1t-q6v1},
  url = {https://link.aps.org/doi/10.1103/pz1t-q6v1}
}

@article{PhysRevLett.132.232503,
  title = {Impact of Two-Body Currents on Magnetic Dipole Moments of Nuclei},
  author = {Miyagi, T. and Cao, X. and Seutin, R. and others},
  journal = {Phys. Rev. Lett.},
  volume = {132},
  issue = {23},
  pages = {232503},
  numpages = {6},
  year = {2024},
  publisher = {American Physical Society},
  doi = {10.1103/PhysRevLett.132.232503},
  url = {https://link.aps.org/doi/10.1103/PhysRevLett.132.232503}
}

@article{PhysRevC.110.034316,
  title = {Large scale shell model calculation for collectivity in nuclei beyond $^{78}\mathrm{Ni}$},
  author = {Chen, N. and Li, J. G. and Li, H. H.},
  journal = {Phys. Rev. C},
  volume = {110},
  issue = {3},
  pages = {034316},
  numpages = {8},
  year = {2024},
  month = {Sep},
  publisher = {American Physical Society},
  doi = {10.1103/PhysRevC.110.034316},
  url = {https://link.aps.org/doi/10.1103/PhysRevC.110.034316}
}

@article{PhysRevLett.118.032502,
  Author = {Stroberg, S. R. and Calci, A. and Hergert, H. and others},
  Doi = {10.1103/PhysRevLett.118.032502},
  Issue = {3},
  Journal = {Phys. Rev. Lett},
  Month = {Jan},
  Numpages = {6},
  Pages = {032502},
  Publisher = {American Physical Society},
  Title = {Nucleus-{D}ependent {V}alence-{S}pace {A}pproach to {N}uclear {S}tructure},
  Url = {https://link.aps.org/doi/10.1103/PhysRevLett.118.032502},
  Volume = {118},
  Year = {2017}
}

@article{Stro19ARNPS,
  author = "Stroberg, S. R. and Bogner, S. K. and Hergert, H. and others",
  title = "{Non-{E}mpirical {I}nteractions for the {N}uclear {S}hell {M}odel: {A}n {U}pdate}",
  journal = "Ann. Rev. Nucl. Part. Sci",
  volume = "69",
  year = "2019",
  pages = "307-362",
  doi = "10.1146/annurev-nucl-101917-021120",
  url = "https://doi.org/10.1146/annurev-nucl-101917-021120"
}

@article{PhysRevC.102.034320,
  title = {Ab initio multishell valence-space Hamiltonians and the island of inversion},
  author = {Miyagi, T. and Stroberg, S. R. and Holt, J. D. and others},
  journal = {Phys. Rev. C},
  volume = {102},
  issue = {3},
  pages = {034320},
  numpages = {10},
  year = {2020},
  month = {Sep},
  publisher = {American Physical Society},
  doi = {10.1103/PhysRevC.102.034320},
  url = {https://link.aps.org/doi/10.1103/PhysRevC.102.034320}
}

@article{PhysRevC.105.014302,
  title = {Converged ab initio calculations of heavy nuclei},
  author = {Miyagi, T. and Stroberg, S. R. and Navr\'atil, P. and others},
  journal = {Phys. Rev. C},
  volume = {105},
  issue = {1},
  pages = {014302},
  numpages = {14},
  year = {2022},
  month = {Jan},
  publisher = {American Physical Society},
  doi = {10.1103/PhysRevC.105.014302},
  url = {https://link.aps.org/doi/10.1103/PhysRevC.105.014302}
}

@article{Hu22Pb208,
    author = "Hu, B. S. and Jiang, W. G. and Miyagi, T. and others",
    title = "{Ab initio predictions link the neutron skin of $^{208}$Pb to nuclear forces}",
    doi = "10.1038/s41567-022-01715-8",
    url ="https://www.nature.com/articles/s41567-022-01715-8",
    journal = "Nat. Phys",
    volume = "18",
    number = "10",
    pages = "1196",
    year = "2022"
}

@article{PhysRevC.83.031301,
  title = {Improved nuclear matter calculations from chiral low-momentum interactions},
  author = {Hebeler, K. and Bogner, S. K. and Furnstahl, R. J. and others},
  journal = {Phys. Rev. C},
  volume = {83},
  issue = {3},
  pages = {031301(R)},
  numpages = {5},
  year = {2011},
  month = {Mar},
  publisher = {American Physical Society},
  doi = {10.1103/PhysRevC.83.031301},
  url = {https://link.aps.org/doi/10.1103/PhysRevC.83.031301}
}

@article{PhysRevC.96.014303,
  title = {Saturation with chiral interactions and consequences for finite nuclei},
  author = {Simonis, J. and Stroberg, S. R. and Hebeler, K. and others},
  journal = {Phys. Rev. C},
  volume = {96},
  issue = {1},
  pages = {014303},
  numpages = {11},
  year = {2017},
  month = {Jul},
  publisher = {American Physical Society},
  doi = {10.1103/PhysRevC.96.014303},
  url = {https://link.aps.org/doi/10.1103/PhysRevC.96.014303}
}

@article{PhysRevC.102.054301,
  title = {Accurate bulk properties of nuclei from ${A}=2$ to $\ensuremath{\infty}$ from potentials with $\mathrm{\ensuremath{\Delta}}$ isobars},
  author = {Jiang, W. G. and Ekstr\"om, A. and Forss\'en, C. and others},
  journal = {Phys. Rev. C},
  volume = {102},
  issue = {5},
  pages = {054301},
  numpages = {8},
  year = {2020},
  month = {Nov},
  publisher = {American Physical Society},
  doi = {10.1103/PhysRevC.102.054301},
  url = {https://link.aps.org/doi/10.1103/PhysRevC.102.054301}
}

@article{PhysRevC.101.014318,
  title = {Novel chiral {H}amiltonian and observables in light and medium-mass nuclei},
  author = {Som\`a, V. and Navr\'atil, P. and Raimondi, F. and others},
  journal = {Phys. Rev. C},
  volume = {101},
  issue = {1},
  pages = {014318},
  numpages = {19},
  year = {2020},
  month = {Jan},
  publisher = {American Physical Society},
  doi = {10.1103/PhysRevC.101.014318},
  url = {https://link.aps.org/doi/10.1103/PhysRevC.101.014318}
}

@article{SHIMIZU2019372,
title = "Thick-restart block {L}anczos method for large-scale shell-model calculations",
journal = "Comput. Phys. Commun",
volume = "244",
pages = "372",
year = "2019",
issn = "0010-4655",
doi = "https://doi.org/10.1016/j.cpc.2019.06.011",
url = "http://www.sciencedirect.com/science/article/pii/S0010465519301985",
author = "Shimizu, N. and Mizusaki, T. and Utsuno, Y. and others"
}

@Article{Morr18Tin,
  author    = {Morris, T. D. and Simonis, J. and Stroberg, S. R. and others},
  title     = {Structure of the Lightest Tin Isotopes},
  journal   = {Phys. Rev. Lett},
  year      = {2018},
  volume    = {120},
  pages     = {152503},
  month     = {Apr},
  doi      = {10.1103/PhysRevLett.120.152503},
  issue     = {15},
  numpages  = {8},
  publisher = {American Physical Society},
  url      = {https://link.aps.org/doi/10.1103/PhysRevLett.120.152503},
}

@article{Holt21Drip,
    author = "Stroberg, S. R. and Holt, J. D. and Schwenk, A. and others",
    title = "Ab Initio Limits of Atomic Nuclei",
    doi = "10.1103/PhysRevLett.126.022501",
    journal = "Phys. Rev. Lett",
    volume = "126",
    number = "2",
    pages = "022501",
    year = "2021"
}

@article{PhysRevC.105.034333,
  title = {Systematics of $E2$ strength in the $sd$ shell with the valence-space in-medium similarity renormalization group},
  author = {Stroberg, S. R. and Henderson, J. and Hackman, G. and Ruotsalainen, P. and Hagen, G. and Holt, J. D.},
  journal = {Phys. Rev. C},
  volume = {105},
  issue = {3},
  pages = {034333},
  numpages = {10},
  year = {2022},
  month = {Mar},
  publisher = {American Physical Society},
  doi = {10.1103/PhysRevC.105.034333},
  url = {https://link.aps.org/doi/10.1103/PhysRevC.105.034333}
}

\clearpage
\section{End Matter}

For the LSSM calculations, two valence spaces were employed to explicitly examine the role of excitations across the $Z=28$ and $N=50$ shell closures. The first assumes an inert $^{78}$Ni core, with a valence space comprising the $\pi(pf_{5/2}g_{9/2})$ and $\nu(sdg_{7/2}h_{11/2})$ orbitals, using the effective interaction constructed in Ref.~\cite{PhysRevC.110.034316}. The second employs a $^{52}$Ca core and an enlarged valence space comprising the full proton $pf$ shell and the $\nu(f_{5/2}p_{1/2}g_{9/2}sd)$ orbitals, with the effective interaction developed in Refs.~\cite{ShellModel-JGL-2023,PhysRevC.111.044308}, thereby allowing nucleon excitations of up to $6p$-$6h$ across the $Z=28$ and $N=50$ shell closures. Effective $g$-factors of $g_s^{\rm eff}=0.7g_s^{\rm free}$ and effective charges of $e_p=1.5e$ and $e_n=0.5e$ were adopted for the magnetic and quadrupole moments, respectively. 
\begin{figure}[H]
\includegraphics[width=0.48\textwidth]{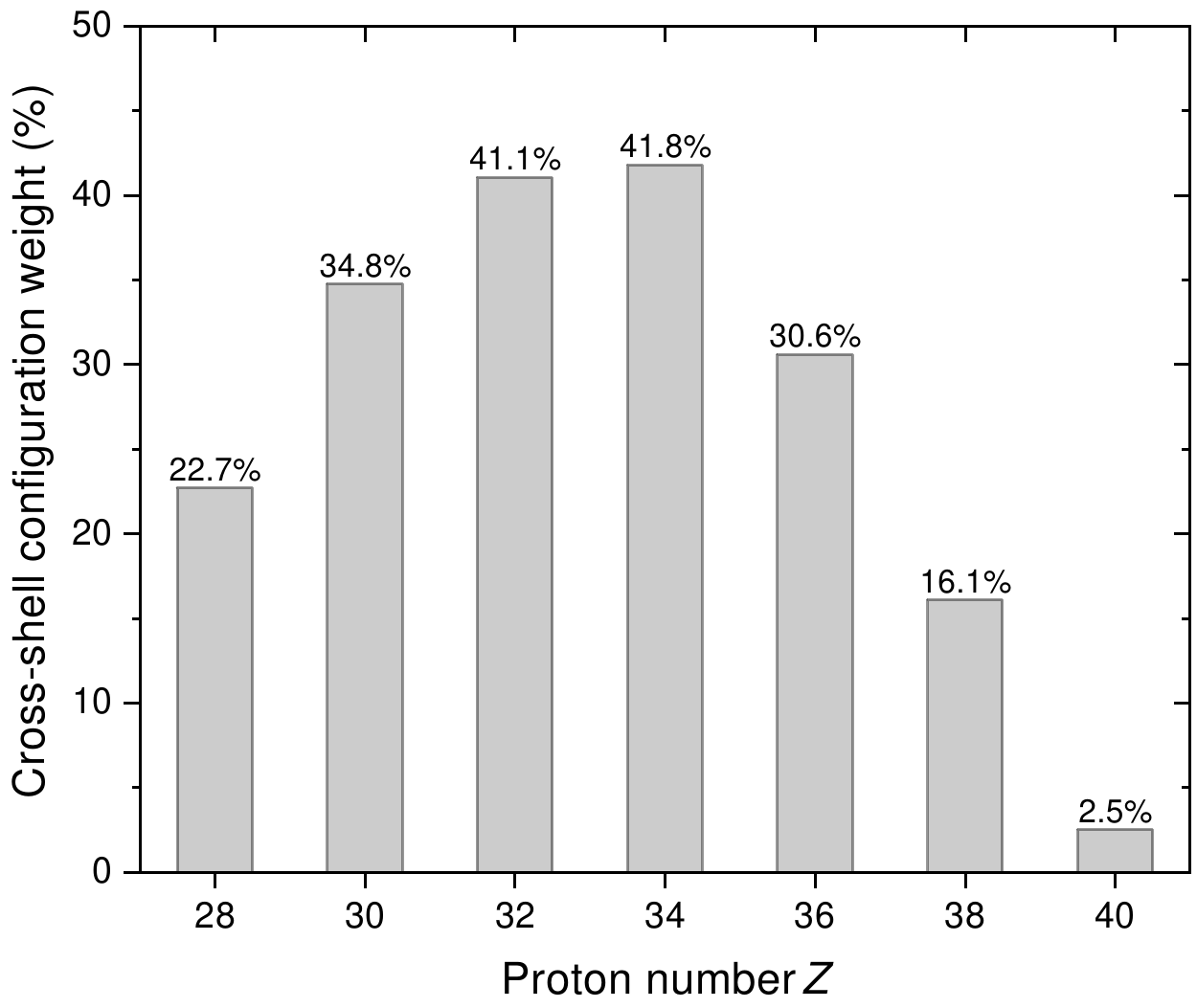}
\vspace{-7mm}
\caption{\footnotesize{Cross-shell configuration weights in the lowest 5/2$^+$ states along the $N=51$ isotonic chain, calculated within the LSSM framework using the $^{52}$Ca-core model space. The weights are summed over the cross-shell configurations involving excitations across the $Z=28$ and/or $N=50$ shell closures.}}
\label{SCE}
\vspace{-3mm}
\end{figure}

To quantify the correlations contained in the enlarged $^{52}$Ca-core calculations, Fig.~\ref{SCE} shows the summed weights of the cross-shell configurations in the lowest 5/2$^+$ states along the $N=51$ isotonic chain, obtained by summing configurations with up to $4p$--$4h$ excitations across the $Z=28$ and/or $N=50$ shell closures.
These weights are extracted directly from the LSSM wave functions. Specifically, each configuration is classified according to the number of protons and/or neutrons excited across the corresponding shell closures. The contributions from configurations with the same particle-hole excitation order are then summed, using the squared wave-function amplitudes as weights.
The total contribution becomes substantial toward the middle of the chain, reaching about 42\% around $Z=32\text{--}34$. For $^{81}$Zn at $Z=30$, it amounts to 34.8\%, providing a microscopic measure of the cross-shell correlations discussed in the main text. Its systematic evolution is consistent with that inferred from the electromagnetic moments in Fig.~\ref{fig:moments}, supporting the interpretation that their deviations from the single-particle picture arise from enhanced cross-shell configuration mixing.

For the \textit{ab initio} VS-IMSRG calculations, three commonly used chiral two- and three-nucleon interactions are employed: 1.8/2.0(EM)~\cite{PhysRevC.83.031301,PhysRevC.96.014303}, $\Delta$NNLO$_{\rm GO}$~\cite{PhysRevC.102.054301}, and N$^3$LO$_{\rm Texas}$~\cite{hu2026-texas}. The 1.8/2.0(EM) is only fitted to $A \le 4$ data, but generally reproduces ground state energies up to the heavy-mass region well ~\cite{Morr18Tin,Holt21Drip,PhysRevC.101.014318}. $\Delta$NNLO$_{\rm GO}$ includes explicit $\rm{\Delta}$-isobar degrees of freedom and optimizes NN and 3N forces simultaneously at the N$^2$LO level. The newly-developed N$^3$LO$_{\rm Texas}$ provides accurate descriptions of binding energies and charge radii in medium- and heavy-mass nuclei, while also reproducing key excited states and nuclear matter properties near saturation~\cite{hu2026-texas}. 
Working in an initial harmonic-oscillator basis of 13 major shells at frequency 16 MeV, an additional cut on 3N matrix elements storage $E_{\rm 3max}$ = 24 \cite{PhysRevC.105.014302} is imposed. The transformation to Hartree-Fock basis is first performed, then the VS-IMSRG~\cite{PhysRevLett.118.032502,Stro19ARNPS,PhysRevC.102.034320} is used to construct approximate unitary transformations to decouple a core and valence-space Hamiltonian from the full $A$-body Hamiltonian. Calculations were performed in valence spaces of $\pi pf_{5/2}g_{9/2}$ and $\nu sdg_{7/2}h_{11/2}$ and $\pi pf$ and $\nu f_{5/2}p_{1/2}g_{9/2}sd$, built on both $^{78}$Ni and $^{52}$Ca cores, respectively, allowing the sensitivity to cross-shell correlations to be assessed. The electromagnetic $M1$ and $E2$ operators were consistently transformed within the VS-IMSRG framework~\cite{PhysRevLett.118.032502, Stro19ARNPS, PhysRevC.102.034320, Hu22Pb208, PhysRevLett.132.232503,Stro22E2}. Note that only bare nucleon charges and free $g$-factors are used here, unlike in the LSSM calculations, and two-body magnetic moment operator contributions have been included as well. Finally, the resulting valence-space Hamiltonians were diagonalized using \textsc{KShell}~\cite{SHIMIZU2019372}.

\end{document}